\documentclass[12pt,thmsa]{article}

\begin{document}

\textbf{Invariant} \textbf{Relativistic\ Electrodynamics. Clifford Algebra
Approach }\bigskip

\qquad Tomislav Ivezi\'{c}

\qquad \textit{Ru%
\mbox
 {\it{d}\hspace{-.15em}\rule[1.25ex]{.2em}{.04ex}\hspace{-.05em}}er Bo\v
{s}kovi\'{c} Institute, P.O.B. 180, 10002 Zagreb, Croatia}

\textit{\qquad ivezic@irb.hr\bigskip }

\qquad In the usual Clifford algebra formulation of electrodynamics the
Faraday bivector field $F$ is decomposed into \emph{the observer dependent }%
sum of a relative vector $\mathbf{E}$ and a relative bivector $e_{5}\mathbf{B%
}$ by making a space-time split, which depends on the observer velocity. ($%
\mathbf{E}$ corresponds to the three-dimensional electric field vector, $%
\mathbf{B}$ corresponds to the three-dimensional magnetic field vector and $%
e_{5}$ is the (grade-4) pseudoscalar.) In this paper it is proved that the
space-time split and the relative vectors are not relativistically correct,
which means that the ordinary Maxwell equations with $\mathbf{E}$ and $%
\mathbf{B}$ and the field equations (FE) with $F$ are not physically
equivalent. Therefore we present \emph{the observer independent }%
decomposition of $F$ by using the 1-vectors of electric $E$ and magnetic $B$
fields. The equivalent, invariant, formulations of relativistic
electrodynamics (independent of the reference frame and of the chosen
coordination for that frame) which use $F,$ $E$ and $B,$ the real
multivector $\Psi =E-e_{5}cB$ and the complex 1-vector $\Psi =E-icB$ are
developed and presented here. \emph{The new observer independent FE }are
presented in formulations with $E$ and $B,$ with real and complex $\Psi $.
When the sources are absent the FE with real and complex $\Psi $ become
Dirac like relativistic wave equations for the free photon. The expressions
for \emph{the observer independent} stress-energy vector $T(v)$ (1-vector)$,$
energy density $U$ (scalar), the Poynting vector $S$ and the momentum
density $g$ (1-vectors), the angular momentum density $M$ (bivector) and the
Lorentz force $K$ (1-vector) are directly derived from the FE. The local
conservation laws are also directly derived from the FE and written in an
invariant way. \bigskip

\noindent \emph{Henceforth space by itself, and time by itself, are doomed }

\noindent \emph{to fade away into mere shadows and only a kind of union of }

\noindent \emph{the two will preserve an independent reality. H. Minkowski}%
\bigskip

\textbf{I. INTRODUCTION} \medskip

In the usual Clifford algebra treatments, e.g. $\left[ 1-3\right] $, of
electrodynamics the Maxwell equations (ME) are written as a single equation
using the electromagnetic field strength $F$ (a bivector) and the gradient
operator $\partial $ (1-vector). (As expressed in $\left[ 3\right] $ (Found.
Phys. \textbf{23}, 1295 (1993)) the reference $\left[ 4\right] $ \textit{%
Clifford Algebra to Geometric Calculus} is: ''one of the most stimulating
modern textbooks of applied mathematics, full of powerful formulas waiting
for physical application.'') In order to get the more familiar form the
field bivector $F$ is expressed in terms of the sum of a relative vector $%
\mathbf{E}$ (corresponds to the three-dimensional electric field vector) and
a relative bivector $e_{5}\mathbf{B}$ ($\mathbf{B}$ corresponds to the
three-dimensional magnetic field vector, and $e_{5}$ is the (grade-4)
pseudoscalar) by making a space-time split, which depends on \emph{the
observer velocity. }It is considered in such formulation that the ME written
in terms of $F$ and of $\mathbf{E}$ and $\mathbf{B}$ are completely
equivalent. The components of $\mathbf{E}$ and $\mathbf{B}$ are considered
to define in a unique way the components of $F$. Moreover in order to get
the wave theory of electromagnetism the vector potential $A$ is introduced
and $F$ is defined in terms of $A$. Thus such formulation with relative
vectors $\mathbf{E,}$ $\mathbf{B}$ and with 1-vector $A$ is not only
observer dependent but also gauge dependent.

However in the recent works $\left[ 5-7\right] $ an invariant formulation of
special relativity (SR)\ is proposed (see also $\left[ 8\right] $) and
compared with different experiments, e.g., the ''muon'' experiment, the
Michelson-Morley type experiments, the Kennedy-Thorndike type experiments
and the Ives-Stilwell type experiments. In such invariant formulation of SR
a physical quantity in the 4D spacetime is mathematically represented either
by a tensor (when no basis has been introduced) or equivalently by a
coordinate-based geometric quantity (CBGQ) comprising both components and a
basis (when some basis has been introduced). \emph{This invariant
formulation is independent of the reference frame and of the chosen
coordination for that frame.} The CBGQs representing some 4D physical
quantity in different relatively moving inertial frames of reference (IFRs),
or in different coordinations of the chosen IFR, are all mathematically
equal and thus they are \emph{the same quantity }for different observers, or
in different coordinations (this fact is the real cause for the name
invariant SR). \emph{It is taken in the invariant SR that such 4D tensor
quantities are well-defined not only mathematically but also experimentally,
as measurable quantities with real physical meaning. }The complete and
well-defined measurement from this invariant SR viewpoint is such
measurement in which all parts of some 4D quantity are measured. The
invariant SR is compared with the usual covariant formulation, which mainly
deals with the basis components of tensors in a specific, i.e., Einstein's
coordination (EC). In the EC the Einstein synchronization $\left[ 9\right] $
of distant clocks and Cartesian space coordinates $x^{i}$ are used in the
chosen IFR. Further the invariant SR is compared with the usual noncovariant
approach to SR in which some quantities are not 4D tensor quantities, but
rather quantities from ''3+1'' space and time, e.g., the synchronously
determined spatial length (the Lorentz contraction) $\left[ 9\right] $. It
is shown in $\left[ 6\right] $ that all the experiments (when they are
complete from the viewpoint of the invariant SR) are in agreement with that
formulation but not always with the usual covariant or noncovariant
approaches to SR. It is also found in $\left[ 5\right] $ that the usual
transformations of the 3D vectors of the electric and magnetic fields $%
\mathbf{E}$ and $\mathbf{B}$ are not relativistically correct.

In this paper it is shown that the space-time split is not relativistically
correct procedure and that the relative vectors are not well-defined
quantities from the SR viewpoint. This means that the ordinary ME with $%
\mathbf{E}$ and $\mathbf{B}$ are not physically equivalent with the observer
independent FE with $F$. Further we write the Lorentz transformations (LT)
in a coordination independent way. Then we present \emph{the observer
independent decomposition of }$F$\emph{\ in terms of 1-vectors }$E$\emph{\
and }$B.$ The new Clifford algebra formulations of relativistic
electrodynamics with 1-vectors $E$ and $B$ and with the real multivector $%
\Psi =E-e_{5}cB,$ or with the complex 1-vector $\Psi =E-icB$ ($i$ is the
unit imaginary), which are completely equivalent to the formulation with the
field bivector $F$, are developed and presented here. The expressions for
the observer independent stress-energy vector $T(v)$ (1-vector)$,$ energy
density $U$ (scalar, i.e., grade-0 multivector), the Poynting vector $S$
(1-vector)$,$ the angular momentum density $M$ (bivector) and the Lorentz
force $K$ (1-vector) are directly derived from the FE and given in all four
formulations. Consequently \emph{the principle of relativity is
automatically included in such formulations with invariant quantities,
whereas in the traditional formulation of SR this principle acts as the
postulate established outside the mathematical formulation of the theory. }%
The local charge-current density and local energy-momentum conservation laws
are derived from the FE. It is also shown that in the real and the complex $%
\Psi $ formulations the FE become a Dirac like relativistic wave equation
for the free photon. The expressions for such geometric 4D quantities are
compared with the familiar ones from the 3D space considering our
definitions in the standard basis $\left\{ \gamma _{\mu }\right\} $ and in
the $\mathcal{R}$ frame (the frame of ''fiducial'' observers) in which $%
E^{0}=B^{0}=0$. This formalism does not make use of the intermediate
electromagnetic 4-potential $A,$ and thus dispenses with the need for the
gauge conditions. The main idea for the whole approach is the same as for
the invariant SR with tensors $\left[ 5-8\right] $, i.e., that \emph{the
physical meaning is attributed, both theoretically and experimentally, only
to the observer independent 4D quantities.} We also remark that the observer
independent quantities introduced here and the FE written in terms of them
are of the same form both in the flat and curved spacetimes. \bigskip

\textbf{II. SHORT\ REVIEW\ OF GEOMETRIC\ ALGEBRA. SPACE-TIME SPLIT. LORENTZ\
TRANSFORMATIONS\bigskip }

\textbf{A. A brief summary of geometric algebra} \medskip

First we provide a brief summary of geometric algebra. We write Clifford
vectors in lower case ($a$) and general multivectors (Clifford aggregate) in
upper case ($A$). The space of multivectors is graded and multivectors
containing elements of a single grade, $r$, are termed homogeneous and
written $A_{r}.$ The geometric (Clifford) product is written by simply
juxtaposing multivectors $AB$. A basic operation on multivectors is the
degree projection $\left\langle A\right\rangle _{r}$ which selects from the
multivector $A$ its $r-$ vector part ($0=$ scalar, $1=$ vector, $2=$
bivector ....). We write the scalar (grade-$0$) part simply as $\left\langle
A\right\rangle .$ The geometric product of a grade-$r$ multivector $A_{r}$
with a grade-$s$ multivector $B_{s}$ decomposes into $A_{r}B_{s}=\left%
\langle AB\right\rangle _{\ r+s}+\left\langle AB\right\rangle _{\
r+s-2}...+\left\langle AB\right\rangle _{\ \left| r-s\right| }.$ The inner
and outer (or exterior) products are the lowest-grade and the highest-grade
terms respectively of the above series $A_{r}\cdot B_{s}\equiv \left\langle
AB\right\rangle _{\ \left| r-s\right| },$ and $A_{r}\wedge B_{s}\equiv
\left\langle AB\right\rangle _{\ r+s}.$ For vectors $a$ and $b$ we have $%
ab=a\cdot b+a\wedge b,$ where $a\cdot b\equiv (1/2)(ab+ba),$ and $a\wedge
b\equiv (1/2)(ab-ba).$ Reversion is an invariant kind of conjugation, which
is defined by $\widetilde{AB}=\widetilde{B}\widetilde{A},$ $\widetilde{a}=a,$
for any vector $a$, and it reverses the order of vectors in any given
expression. Also we shall need the operation called the complex reversion
(for example, when working with complex 1-vector $\Psi =E-ciB$). The complex
reversion of, e.g., $\Psi $, is denoted by an overbar $\overline{\Psi .}$ It
takes the complex conjugate of the scalar (complex) coefficient of each of
the 16 elements in the algebra, and reverses the order of multiplication of
vectors in each multivector.\bigskip

\textbf{B. Standard basis, non-standard bases, and the space-time
split\medskip }

In the treatments, e.g., $\left[ 1-3\right] $, one usualy introduces the
standard basis. The generators of the spacetime algebra (STA) (the Clifford
algebra generated by Minkowski spacetime) are taken to be four basis vectors
$\left\{ \gamma _{\mu }\right\} ,\mu =0...3,$ satisfying $\gamma _{\mu
}\cdot \gamma _{\nu }=\eta _{\mu \nu }=diag(+---).$ This basis is a
right-handed orthonormal frame of vectors in the Minkowski spacetime $M^{4}$
with $\gamma _{0}$ in the forward light cone. The $\gamma _{k}$ ($k=1,2,3$)
are spacelike vectors. This algebra is often called the Dirac algebra $D$
and the elements of $D$ are called $d-$numbers. The $\gamma _{\mu }$
generate by multiplication a complete basis, the standard basis, for STA: $%
1,\gamma _{\mu },\gamma _{\mu }\wedge \gamma _{\nu },\gamma _{\mu }\gamma
_{5,}\gamma _{5}$ ($2^{4}=16$ independent elements). $\gamma _{5}$ is the
pseudoscalar for the frame $\left\{ \gamma _{\mu }\right\} .$

We remark that the standard basis corresponds, in fact, to the specfic
coordination, i.e., the EC, of the chosen IFR. However different
coordinations of an IFR are allowed and they are all equivalent in the
description of physical phenomena. For example, in $\left[ 5\right] $ two
very different, but completely equivalent coordinations, the EC and
''radio'' (''r'') coordination, are exposed and exploited throughout the
paper. For more detail about the ''r'' coordination see, e.g., $\left[
5\right] $ and references therein.

The next step in the usual treatments, e.g., $\left[ 1-3\right] $, is the
introduction of a space-time split and the relative vectors. Since the usual
STA deals exclusively with the EC it is possible to say that a given IFR is
completely characterized by a single future-pointing, timelike unit vector $%
\gamma _{0}$ ($\gamma _{0}$ is tangent to the world line of an observer at
rest in the $\gamma _{0}$-system). By singling out a particular time-like
direction $\gamma _{0}$ we can get a unique mapping of spacetime into the
even subalgebra of STA (the Pauli subalgebra). For each spacetime point (or
event) $x$ this mapping is specified by
\begin{equation}
x\gamma _{0}=ct+\mathbf{x,\quad }ct=x\cdot \gamma _{0},\ \mathbf{x}=x\wedge
\gamma _{0}.  \label{split}
\end{equation}
To each event $x$ the equation (\ref{split}) assigns a unique time and
position in the $\gamma _{0}$-system. The set of all position vectors $%
\mathbf{x}$ is the 3-dimensional position space of the observer $\gamma _{0}$
and it is designated by $P^{3}=P^{3}(\gamma _{0})=\left\{ \mathbf{x}=x\wedge
\gamma _{0}\right\} .$ The elements of $P^{3}$ are all spacetime bivectors
with $\gamma _{0}$ as a common factor ($x\wedge \gamma _{0}).$ They are
called \textit{the relative vectors} (relative to $\gamma _{0})$ and they
will be designated in boldface. Then a standard basis $\left\{ \mathbf{%
\sigma }_{k};k=1,2,3\right\} $ for $P^{3},$ which corresponds to a standard
basis $\left\{ \gamma _{\mu }\right\} $ for $M^{4}$ is given as $\mathbf{%
\sigma }_{k}=\gamma _{k}\wedge \gamma _{0}=\gamma _{k}\gamma _{0}.$ The
invariant distance $x^{2}$ then decomposes as $x^{2}=(x\gamma _{0})(\gamma
_{0}x)=(ct-\mathbf{x})(ct+\mathbf{x})=c^{2}t^{2}-\mathbf{x}^{2}.$ The
explicit appearance of $\gamma _{0}$ in (\ref{split}) imply that \emph{the
space-time split is observer dependent}, i.e., it is dependent on the chosen
IFR. It has to be noted that in the EC the space-time split of the position
1-vector $x$ (\ref{split}) gives separately the space and time components of
$x$ with their usual meaning, i.e., as in the prerelativistic physics, and
(as shown above) in the invariant distance $x^{2}$ the spatial and temporal
parts are also separated. (In the ''r'' coordination there is no space-time
split and also in $x^{2}$ the spatial and temporal parts are not separated,
see $\left[ 5\right] $.) This does not mean that the EC does have some
advantage relative to other coordinations and that the quantities in the EC
are more physical than, e.g., those in the ''r'' coordination.

Different coordinations refer to the same IFR, say the $S$ frame. But if we
consider the geometric quantity, the position 1-vector, $x$ in another
relatively moving IFR $S^{\prime },$ which is characterized by $\gamma
_{0}^{\prime },$ then the space-time split in $S^{\prime }$ and in the EC is
$x\gamma _{0}^{\prime }=ct^{\prime }+\mathbf{x}^{\prime }\mathbf{,}$ and
this $x\gamma _{0}^{\prime }$ is not obtained by the LT (or any other
coordinate transformations) from $x\gamma _{0}.$ (The hypersurface $%
t^{\prime }=const.$ is not connected in any way with the hypersurface $%
t=const.$) Thus the customary Clifford algebra approaches to SR start with
the geometric, i.e., coordinate-free, quantities, e.g., $x,x^{2},$ etc.$,$
which are physically well-defined. However the use of the space-time split
introduces in the customary approaches such coordinate-dependent quantities
which are not physically well-defined since they cannot be connected by the
LT. The main difference between our invariant approach to SR (by the use of
the Clifford algebra) and the other Clifford algebra approaches is that in
our approach, as already said, \emph{the physical meaning is attributed,
both theoretically and experimentally, only to the geometric 4D quantities,
and not to their parts. Thus there is no need and moreover it is not
physical from the viewpoint of the invariant SR to introduce the space-time
split of the geometric 4D quantity.} We consider, in the same way as H.
Minkowski (the motto in this paper), that the spatial and the temporal
components (e.g., $\mathbf{x}$ and $t,$ respectively) of some geometric 4D
quantity (e.g., $x$) are not physically well-defined quantities. Only their
union is physically well-defined and only such quantity does have an
independent reality.

Thus instead of the standard basis $\left\{ \gamma _{\mu }\right\} ,$ $\mu
=0...3,$ for $M^{4}$ we can use some basis $\left\{ e_{\mu }\right\} $ (the
metric tensor of $M^{4}$ is then defined as $g_{\mu \nu }=e_{\mu }\cdot
e_{\nu }$) and its dual basis $\left\{ e^{\mu }\right\} ,$ where the set of
base vectors $e^{\mu }$ is related to the $e_{\mu }$ by the conditions $%
e_{\mu }\cdot e^{\nu }=\delta _{\mu }^{\nu }$. The pseudoscalar $e_{5}$ of a
frame $\left\{ e_{\mu }\right\} $ is defined by $e_{5}=e_{0}\wedge
e_{1}\wedge \wedge e_{2}\wedge e_{3}.$ Then, e.g., the position 1-vector $x$
can be decomposed in the $S$ and $S^{\prime }$ frames and in the standard
basis $\left\{ \gamma _{\mu }\right\} $ and some non-standard basis $\left\{
e_{\mu }\right\} $ as $x=x^{\mu }\gamma _{\mu }=x^{\mu ^{\prime }}\gamma
_{\mu ^{\prime }}=....=x_{e}^{\mu ^{\prime }}e_{\mu ^{\prime }}.$ The primed
quantities are the Lorentz transforms of the unprimed ones. Similarly any
multivector $A$ can be written as an invariant quantity with the components
and the basis, i.e., as the CBGQ. In such interpretation the LT are
considered as passive transformations; both the components and the base
vectors are transformed but the whole geometric quantity remains unchanged.
Thus we see that under the passive LT a well-defined quantity on the 4D
spacetime, i.e., a CBGQ, is an invariant quantity. This doesn't hold for the
relative vectors and thence they are not well-defined quantities from the SR
viewpoint.\bigskip

\textbf{C. Lorentz transformations\medskip }

In the usual Clifford algebra formalism, e.g., $\left[ 1-4\right] $, the LT
are considered as active transformations; the components of, e.g., some
1-vector relative to a given IFR (with the standard basis $\left\{ \gamma
_{\mu }\right\} $) are transformed into the components of a new 1-vector
relative to the same frame (the basis $\left\{ \gamma _{\mu }\right\} $ is
not changed). Furthermore the LT are described with rotors $R,$ $R\widetilde{%
R}=1,$ in the usual way as $p\rightarrow p^{\prime }=Rp\widetilde{R}=p_{\mu
^{\prime }}\gamma ^{\mu }.$ But every rotor in spacetime can be written in
terms of a bivector as $R=e^{\theta /2}.$ For boosts in arbitrary direction $%
e^{\theta /2}=(1+\gamma +\gamma \beta \gamma _{0}n)/(2(1+\gamma ))^{1/2},$ $%
\theta =\alpha \gamma _{0}n,$ $\beta $ is the scalar velocity in units of $c$%
, $\gamma =(1-\beta ^{2})^{-1/2}$, or in terms of an `angle' $\alpha $ we
have $\tanh \alpha =\beta ,$ $\cosh \alpha =\gamma ,$ $\sinh \alpha =\beta
\gamma ,$ and $n$ is not the basis vector but any unit space-like vector
orthogonal to $\gamma _{0};$ $e^{\theta }=\cosh \alpha +\gamma _{0}n\sinh
\alpha .$ (One can also express the relationship between the two relatively
moving frames $S$ and $S^{\prime }$ in terms of rotor as $\gamma _{\mu
^{\prime }}=R\gamma _{\mu }\widetilde{R}.$) The above explicit form for $%
R=e^{\theta /2}$ is frame independent but it is coordination dependent since
it refers to the EC.

Here a coordination independent form for the LT is introduced and it can be
used both in an active way (when there is no basis) or in a passive way
(when some basis is introduced). The main step is the introduction of the
1-vector $u=cn,$ which represents \emph{the proper velocity of the frame} $S$
\emph{with respect to itself}. Then taking that $v$ is 1-vector of the
velocity of $S^{\prime }$ relative to $S$ we write the component form of $L$
in some basis $\left\{ e_{\mu }\right\} $ which$,$ as already said, does not
need to be the standard basis, as
\begin{eqnarray}
L_{\nu }^{\mu } &=&g_{\nu }^{\mu }+2u^{\mu }v_{\nu }c^{-2}-  \nonumber \\
&&(u^{\mu }+v^{\mu })(u_{\nu }+v_{\nu })/c^{2}(1+u\cdot v/c^{2}),  \label{L}
\end{eqnarray}
or with the components and the basis, i.e., as the CBGQ, $L=L_{\nu }^{\mu
}e_{\mu }e^{\nu },$ see the second paper in $\left[ 8\right] $ and $\left[
5\right] ;$ actually this form of the LT is a generalization to arbitrary
coordination of the covariant form of the LT in the EC given in $\left[
10\right] $. The rotor connected with such $L$ is
\begin{equation}
R=L/(\widetilde{L}L)^{1/2}=L_{\nu }^{\mu }e_{\mu }e^{\nu }/(\widetilde{L}%
L)^{1/2},\quad \widetilde{L}L=8(\gamma +1),\quad \gamma =u\cdot v/c^{2},
\label{R1}
\end{equation}
It can be also written as
\begin{eqnarray}
R &=&\left\langle R\right\rangle +\left\langle R\right\rangle _{2}=\cosh
\alpha /2+((u\wedge v)/\left| u\wedge v\right| )\sinh \alpha /2=  \nonumber
\\
&&\exp ((\alpha /2)(u\wedge v)/\left| u\wedge v\right| ),  \label{R2} \\
R &=&((1+u\cdot v)/2)^{1/2}+((u\wedge v)/\left| u\wedge v\right|
)((-1+u\cdot v)/2)^{1/2}.  \nonumber
\end{eqnarray}
One also can solve $L_{\nu }^{\mu }$ in terms of $R$ as
\begin{equation}
L_{\nu }^{\mu }=\left\langle e_{\nu }\widetilde{R}e^{\mu }R\right\rangle .
\label{RL}
\end{equation}
The usual results are recovered when the standard basis $\left\{ \gamma
_{\mu }\right\} ,$ i.e., the EC is used. But these results for $L$ and $R$
hold also for other bases, i.e., coordinations. (Thus one can easily find
the LT in the ''r'' coordination, $L_{\nu ,r}^{\mu },$ and compare it with
the corresponding result in $\left[ 5\right] $.)\bigskip

\textbf{III.\ THE\ }$F$\ \textbf{FORMULATION\ OF\ ELECTRODYNAMICS AND THE
PROOF THAT\ THE\ SPACE-TIME\ SPLIT\ AND\ THE\ TRANSFORMATIONS\ OF\ RELATIVE\
VECTORS\ }$\mathbf{E}$\textbf{\ AND }$\mathbf{B}$\textbf{\ ARE\ NOT\
RELATIVISTICALLY\ CORRECT \bigskip }

\textbf{A.} \textbf{The\ }$F$\ \textbf{formulation\ of\ electrodynamics
\medskip }

We start the exposition of electrodynamics writing the FE in terms of $F$ $%
\left[ 1-3\right] $; an electromagnetic field is represented by a
bivector-valued function $F=F(x)$ on spacetime. The source of the field is
the electromagnetic current $j$ which is a 1-vector field. Then using that
the gradient operator $\partial $ is a 1-vector FE can be written as a
single equation
\begin{equation}
\partial F=j/\varepsilon _{0}c,\quad \partial \cdot F+\partial \wedge
F=j/\varepsilon _{0}c.  \label{MEF}
\end{equation}
The trivector part is identically zero in the absence of magnetic charge.
Notice that in $\left[ 1-3\right] $ the FE (\ref{MEF}) are considered to
encode all of the ME, i.e., that the FE (\ref{MEF}) and the usual ME with $%
\mathbf{E}$ and $\mathbf{B}$ are physically equivalent. Our discussion will
show that this is not true.

The field bivector $F$ yields the complete description of the
electromagnetic field and, in fact, there is no need to introduce either the
field vectors or the potentials. For the given sources the Clifford algebra
formalism enables one to find in a simple way the electromagnetic field $F.$
Namely the gradient operator $\partial $ is invertible and (\ref{MEF}) can
be solved for $F=\partial ^{-1}(j/\varepsilon _{0}c),$ see, e.g., $\left[
1-3\right] .$

In the Clifford algebra formalism one can easily derive the expressions for
the stress-energy vector $T(v)$ and the Lorentz force $K$ directly from FE (%
\ref{MEF}) and from the equation for $\widetilde{F},$ the reverse of $F,$ $%
\widetilde{F}\widetilde{\partial }=\widetilde{j}/\varepsilon _{0}c$ ($%
\widetilde{\partial }$ differentiates to the left instead of to the right).
Indeed, using (\ref{MEF}) and from the equation for $\widetilde{F}$ one finds

\begin{equation}
T(\partial )=(-\varepsilon _{0}/2)(F\partial F)=j\cdot F/c=-K,  \label{TEF}
\end{equation}
where in $(F\partial F)$ the derivative $\partial $ operates to the left and
to the right by the chain rule. The stress-energy vector $T(v)$ $\left[
1-3\right] $ for the electromagnetic field is then defined in the $F$
formulation as
\begin{equation}
T(v)=T(v(x),x)=-(\varepsilon _{0}/2c)\left\langle FvF\right\rangle _{1}.
\label{ten}
\end{equation}
We note that $T(v)$ is a vector-valued linear function on the tangent space
at each spacetime point $x$ describing the flow of energy-momentum through a
surface with normal $n=n(x);$ $v=cn.$

The right hand side of (\ref{TEF}) yields the expression for the Lorentz
force $K,$ $K=F\cdot j/c.$ This relation shows that the Lorentz force $K$
can be interpreted as the rate of energy-momentum transfer from the source $%
j $ to the field $F$. The Lorentz force in the $F$ formulation for a charge $%
q$ is $K=(q/c)F\cdot u,$ where $u$ is the velocity 1-vector of a charge $q$
(it is defined to be the tangent to its world line).

The stress-energy vector $T(v)$ can be written in the following form
\begin{equation}
T(v)=-(\varepsilon _{0}/2c)\left[ (F\cdot F)v+2(F\cdot v)\cdot F\right] .
\label{ten1}
\end{equation}
We write $T(v)$ (\ref{ten1}) in a \emph{new form} as a sum of $v$-parallel
part ($v-\parallel $) and $v$-orthogonal part ($v-\perp $)
\begin{eqnarray}
T(v) &=&-(\varepsilon _{0}/2c)\left[ -(F\cdot F)+(2/c^{2})(F\wedge
v)^{2}\right] v+  \nonumber \\
&&-(\varepsilon _{0}/c)\left[ (F\cdot v)\cdot F-(1/c^{2})(F\cdot
v)^{2}v\right] .  \label{ste}
\end{eqnarray}
The first term in (\ref{ste}) is $v-\parallel $ part and it yields the
energy density $U.$ Namely using $T(v)$ and the fact that $v\cdot T(v)$ is
positive for any timelike vector $v$ we construct the expression for \emph{%
the observer independent energy density} $U$ contained in an electromagnetic
field as $U=v\cdot T(v)/c=(1/c)\left\langle vT(v)\right\rangle ,$ (scalar,
i.e., grade-0 multivector). Thus in terms of $F$ and (\ref{ste}) $U$ becomes
\begin{equation}
U=(-\varepsilon _{0}/2c^{2})\left\langle FvFv\right\rangle =(\varepsilon
_{0}/2)\left[ (F\cdot F)-(2/c^{2})(F\wedge v)^{2}\right] .  \label{uen1}
\end{equation}
The second term in (\ref{ste}) is $v-\perp $ part and it is $(1/c)S$, where $%
S$ is \emph{the observer independent expression for the Poynting vector }%
(1-vector),
\begin{equation}
S=-\varepsilon _{0}\left[ (F\cdot v)\cdot F-(1/c^{2})(F\cdot v)^{2}v\right] ,
\label{po1}
\end{equation}
and, as can be seen, $v\cdot S=0$. Thus $T(v)$ expressed by $U$ and $S$ is
\begin{equation}
T(v)=(1/c)(Uv+S).  \label{uis}
\end{equation}
\emph{The observer independent momentum density} $g$ is defined as $%
g=(1/c^{2})S$, i.e., $g$ is $(1/c)$ of the $v-\perp $ part from (\ref{ste})
\begin{equation}
g=-(\varepsilon _{0}/c^{2})\left[ (F\cdot v)\cdot F-(1/c^{2})(F\cdot
v)^{2}v\right] .  \label{ge}
\end{equation}
From $T(v)$ (\ref{ste}) one finds also the expression for \emph{the observer
independent angular-momentum density} $M$
\begin{equation}
M=(1/c)T(v)\wedge x=(U/c^{2})v\wedge x+g\wedge x.  \label{em}
\end{equation}
The first term in (\ref{em}) corresponds to the ''orbital'' angular-momentum
density whereas the second term yields the ''spin'' or intrinsic
angular-momentum density. It has to be emphasized once again that all these
definitions are the definitions of invariant quantities, i.e., frame and
coordination independent quantities.

All these quantities can be written in some basis $\left\{ e_{\mu }\right\}
, $ which does not need to be the standard basis, as CBGQs. Thus $T(v)$ (\ref
{ten1}) becomes
\begin{equation}
T(v)=-(\varepsilon _{0}/2c)\left[ (-1/2)F^{\alpha \beta }F_{\alpha \beta
}v^{\rho }e_{\rho }+2F^{\alpha \beta }F_{\alpha \rho }v^{\rho }e_{\beta
}\right] ,  \label{ten2}
\end{equation}
the energy density $U$ (\ref{uen1}) is
\begin{equation}
U=(\varepsilon _{0}/2)\left[ (1/2)F^{\alpha \beta }F_{\alpha \beta
}-(2/c^{2})F^{\alpha \beta }F_{\alpha \rho }v^{\rho }v_{\beta }\right] ,
\label{un2}
\end{equation}
and the Poynting vector $S$ (\ref{po1}) becomes
\begin{equation}
S=-\varepsilon _{0}\left[ F^{\alpha \beta }F_{\alpha \rho }v^{\rho }e_{\beta
}-(1/c^{2})F^{\alpha \beta }F_{\alpha \rho }v^{\rho }v_{\beta }v^{\lambda
}e_{\lambda }\right] .  \label{po2}
\end{equation}
(The energy-momentum tensor $T^{\mu \nu }$ in the $F$ (and the $E,B$)
formulations will be presented in Sec. IV C.) \bigskip

\textbf{B. The\ local conservation laws in the }$F$\textbf{- formulation}
\textbf{\medskip }

It is well-known that from the FE in the $F$- formulation (\ref{MEF}) one
can derive a set of conserved currents. Of course the same holds for all
other formulations; the $E,B$-formulation, the real and the complex $\Psi $%
-formulation, which will be discussed below. Thus, for example, in the $F$-
formulation one derives in the standard way that $j$ from (\ref{MEF}) is a
conserved current. Simply, the vector derivative $\partial $ is applied to
the FE (\ref{MEF}) which yields
\[
(1/\varepsilon _{0}c)\partial \cdot j=\partial \cdot (\partial \cdot F).
\]
Using the identity $\partial \cdot (\partial \cdot M(x))\equiv 0$ ($M(x)$ is
a multivector field) one obtains \emph{the local charge conservation law}
\begin{equation}
\partial \cdot j=0.  \label{cjo}
\end{equation}

In a like manner we find from (\ref{TEF}) that
\begin{equation}
\partial \cdot T(v)=0  \label{coti}
\end{equation}
for the free fields. This is a \emph{local energy-momentum conservation law}%
. In the derivation of (\ref{TEF}) we used the fact that $T(v)$ is
symmetric, i.e., that $a\cdot T(b)=T(a)\cdot b.$ Namely using overdots the
expression for $T(\partial )$ ($T(\partial )=(-\varepsilon _{0}/2)(F\partial
F),$ where $\partial $ operates to the left and to the right by the chain
rule) can be written as $T(\partial )=\stackrel{\cdot }{T}(\stackrel{\cdot }{%
\partial })=(-\varepsilon _{0}/2)(\stackrel{\cdot }{F}\stackrel{\cdot }{%
\partial }F+F\stackrel{\cdot }{\partial }\stackrel{\cdot }{F})=0,$ since in
the absence of sources $\partial F=\stackrel{\cdot }{F}\stackrel{\cdot }{%
\partial }=0$ (the overdot denotes the multivector on which the derivative
acts). Then from the above mentioned symmetry of $T$ one finds that $%
\stackrel{\cdot }{T}(\stackrel{\cdot }{\partial })\cdot v=\partial \cdot
T(v)=0$, $\forall \ const.\ v$, which proves the equation (\ref{coti}).

Inserting the expression (\ref{uis}) for $T(v)$ into the local
energy-momentum conservation law (\ref{coti}) we find
\begin{equation}
(v\cdot \partial )U+\partial \cdot S=0.  \label{Poy}
\end{equation}
The relation (\ref{Poy}) is the well-known Poynting's theorem but now
completely written in terms of the observer independent quantities. Let us
introduce the standard basis $\left\{ \gamma _{\mu }\right\} ,$ i.e., an IFR
with the EC, and in the $\left\{ \gamma _{\mu }\right\} $ basis we choose
that $v=c\gamma _{0}$, or in the component form it is $v^{\mu }(c,0,0,0).$
Then the familiar form of Poynting's theorem is recovered in such coordinate
system
\begin{equation}
\partial U/\partial t+\partial _{i}S^{i}=0,\qquad i=1,2,3.  \label{Poy1}
\end{equation}
It is worthwhile to note that although $U$ (\ref{uen1}) and $S$ (\ref{ste}),
taken separately, are well-defined observer independent quantities, the
relations (\ref{uis}), (\ref{coti}) and (\ref{Poy}) reveal that only $T(v)$ (%
\ref{uis}), as a whole quantity, i.e., the combination of $U$ and $S,$
enters into a fundamental physical law, the local energy-momentum
conservation law (\ref{coti}). Thence one can say that only $T(v)$ (\ref{uis}%
), as a whole quantity, does have a real physical meaning, or, better to
say, a physically correct interpretation. An interesting example that
emphasizes this point is the case of a uniformly accelerated charge. In the
usual (3D) approach to the electrodynamics ($\left[ 11\right] $ Sec. 6.8.)
the Poynting vector $S$ is interpreted as an energy flux due to the
propagation of fields. In such an interpretation it is not clear how the
fields propagate along the axis of motion since for the field points on the
axis of motion one finds that $S=0$ (there is no energy flow) but at the
same time $U\neq 0$ (there is an energy density). Our approach reveals that
the important quantity is $T(v)$ and not $S$ and $U$ taken separately. $T(v)$
is $\neq 0$ everywhere on the axis of motion and the local energy-momentum
conservation law (\ref{coti}) holds everywhere.

Of course the same law (\ref{coti}) will be obtained in other formulations, $%
E$ and $B$, real and complex $\Psi $ formulations, as well. In the same way
one can derive the local angular momentum conservation law, see $\left[
1\right] ,$ Space-Time Calculus.\bigskip

\textbf{C. The\ space-time\ split\ and\ the\ relative\ vectors\ }$\mathbf{E}$%
\textbf{\ and }$\mathbf{B}$\textbf{\ \medskip }

As already said in the usual Clifford algebra treatments of the
electromagnetism the field bivector $F$ is expressed in terms of the sum of
a relative vector $\mathbf{E}$ and a relative bivector $\gamma _{5}\mathbf{B}
$ by making a space-time split in the $\gamma _{0}$ frame
\begin{eqnarray}
F &=&\mathbf{E}+c\gamma _{5}\mathbf{B,\quad E}=(F\cdot \gamma _{0})\gamma
_{0}=(1/2)(F-\gamma _{0}F\gamma _{0}),  \nonumber \\
\gamma _{5}\mathbf{B} &=&(F\wedge \gamma _{0})\gamma _{0}=(1/2c)(F+\gamma
_{0}F\gamma _{0}).  \label{FB}
\end{eqnarray}
$F$ can be written as the CBGQ in the standard basis $\left\{ \gamma _{\mu
}\right\} $ as
\begin{equation}
F=(1/2)F^{\mu \nu }\gamma _{\mu }\wedge \gamma _{\nu }=F^{0k}\gamma
_{0}\wedge \gamma _{k}+(1/2)F^{kl}\gamma _{k}\wedge \gamma _{l},\quad
k,l=1,2,3.  \label{EF}
\end{equation}
From (\ref{EF}) and (\ref{FB}) one concludes that the relative vectors $%
\mathbf{E}$ and $\mathbf{B}$ are expressed in the standard basis $\left\{
\gamma _{\mu }\right\} $ as
\begin{equation}
\mathbf{E=}F^{0k}\gamma _{0}\wedge \gamma _{k},\quad \gamma _{5}\mathbf{B=}%
(1/2c)F^{kl}\gamma _{k}\wedge \gamma _{l}.  \label{EiB}
\end{equation}
We see from (\ref{EF}) and (\ref{EiB}) that the components of $F$ in the $%
\left\{ \gamma _{\mu }\right\} $ basis give rise to the tensor $F^{\mu \nu
}=\gamma ^{\nu }\cdot (\gamma ^{\mu }\cdot F)=(\gamma ^{\nu }\wedge \gamma
^{\mu })\cdot F,$ which, written out as a matrix, has entries $E_{i}=-F^{0i}$
and $B_{i}=-(1/2)\varepsilon _{ijk}F^{jk}.$ (We write the components $E_{i}$
and $B_{i}$ (and $\varepsilon _{ijk}$) with lowered (generic) subscripts,
since they are not the spatial components of the well-defined quantities on
the 4D spacetime.) It is considered in such formulation that the FE written
in terms of $F$ and the ME with $\mathbf{E}$ and $\mathbf{B}$ (taken in the $%
\left\{ \gamma _{\mu }\right\} $ basis) are completely equivalent. Such
usual interpretation is physically meaningless since $F$ is the well-defined
geometric quantity in the 4D spacetime with the correct transformation
properties while it is not the case for $\mathbf{E}$ and $\mathbf{B}$, as
will be shown below. We consider, in accordance with Minkowski's assertion
(the motto here), that physically meaningful is only the whole geometric
quantity, e.g., the field bivector $F,$ the position 1-vector $x,$ etc.
(when there is no basis) or the corresponding CBGQ $(1/2)F^{\mu \nu }e_{\mu
}\wedge e_{\nu },$ $x^{\mu }e_{\mu },$ (when some basis $\left\{ e_{\mu
}\right\} $ is chosen), and not some parts of it, taken in the specific
representation, e.g., $ct$ and $\mathbf{x}$, or $\mathbf{E}$ and $\mathbf{B}$%
, etc. Such parts of a geometric quantity have no definite physical sense
since they do not transform properly under the LT. Namely the active LT
transform \emph{all components together} leaving the basis unchanged, while
the passive LT transform both, \emph{all components and all basis vectors
together} leaving the whole quantity unchanged. Hence the space-time split
(in the EC) of any geometric quantity defined on the 4D spacetime is, in
fact, an incorrect procedure from the point of view of SR. Also it cannot be
said, as usually argued both in the tensor formalism and in the Clifford
algebra formalism, that $F^{\mu \nu }$ (and thus $F$ too) are determined by
the components of 3D quantities $\mathbf{E}$ and $\mathbf{B}$. $F$ is the
geometric 4D quantity and for the given sources it is completely determined
in an observer independent way by the equation $F=\partial
^{-1}(j/\varepsilon _{0}c)$. We repeat once again that \emph{in the 4D
spacetime the physical meaning is attributed only to the whole 4D geometric
quantity not to its spatial and temporal parts. }This is the main difference
between our approach and others, e.g., $\left[ 1-3\right] $.\bigskip

\textbf{D.} \textbf{The proof that\ the\ transformations\ of\ relative\
vectors\ }$\mathbf{E}$\textbf{\ and }$\mathbf{B}$\textbf{\ are\ not\
relativistically\ correct\medskip }

Let us now explicitly show that the above decomposition of $F$ (\ref{FB}) is
not relativistically correct and that the usual transformations of $\mathbf{E%
}$ and $\mathbf{B}$ are not the LT of quantities that are well-defined on
the 4D spacetime. It can be easily seen that the LT (the active ones) of the
field bivector $F,$ $F^{\prime }=RF\widetilde{R},$ with $\theta =\alpha
\gamma _{0}\gamma _{1}$ (all in the standard basis $\left\{ \gamma _{\mu
}\right\} $), yields
\begin{eqnarray}
F^{\prime } &=&((1+\gamma )/2)[F^{0k}\gamma _{0}\wedge \gamma
_{k}+(1/2)F^{kl}\gamma _{k}\wedge \gamma _{l}]+\gamma \beta F^{0k}\gamma
_{k}\wedge \gamma _{1}+  \nonumber \\
&&\gamma \beta \gamma _{0}[\gamma _{1}(1/2)F^{kl}\gamma _{k}\wedge \gamma
_{l}-(1/2)F^{kl}\gamma _{k}\wedge \gamma _{l}\gamma _{1}]-  \label{FC} \\
&&(\gamma ^{2}\beta ^{2}/2(1+\gamma ))[\gamma _{0}\gamma _{1}F^{0k}\gamma
_{0}\wedge \gamma _{k}\gamma _{0}\gamma _{1}+\gamma _{0}\gamma
_{1}(1/2)F^{kl}\gamma _{k}\wedge \gamma _{l}\gamma _{0}\gamma _{1}].
\nonumber
\end{eqnarray}
Using (\ref{EiB}) one can write (\ref{FC}) as
\begin{eqnarray}
F^{\prime } &=&((1+\gamma )/2)(\mathbf{E}+c\gamma _{5}\mathbf{B})+\gamma
\beta F^{0k}\gamma _{k}\wedge \gamma _{1}+\gamma \beta \gamma _{0}(\gamma
_{1}\cdot c\gamma _{5}\mathbf{B})\mathbf{-}  \nonumber \\
&&((1-\gamma )/2)\gamma _{0}\gamma _{1}(\mathbf{E}+c\gamma _{5}\mathbf{B}%
)\gamma _{0}\gamma _{1}.  \label{FC2}
\end{eqnarray}
The relations (\ref{FC}) and (\ref{FC2}) clearly show that \emph{the active
LT do not transform }$F$ \emph{(\ref{EF}) }into $F^{0^{\prime }k^{\prime
}}\gamma _{0}\wedge \gamma _{k}+(1/2)F^{k^{\prime }l^{\prime }}\gamma
_{k}\wedge \gamma _{l}$ \emph{but introduce some additional terms (e.g.,} $%
\gamma \beta F^{0k}\gamma _{k}\wedge \gamma _{1}$\emph{)}. \emph{The
separated parts} ($\mathbf{E}$ \emph{and} $\mathbf{B}$) \emph{of a
well-defined 4D quantity} $F$ \emph{do not transform into the corresponding
transformed parts} ($F^{0^{\prime }k^{\prime }}\gamma _{0}\wedge \gamma _{k}$
\emph{and} $(1/2)F^{k^{\prime }l^{\prime }}\gamma _{k}\wedge \gamma _{l}$)
\emph{Such result implies that the relative vectors }$\mathbf{E}$\emph{\ and
}$\mathbf{B}$\emph{\ are not well-defined quantities on the 4D spacetime,
since they do not have the correct transformation properties. }Hence it is
not possible to write, as usually supposed, that the transformed $F^{\prime
} $ in the $\gamma _{0}$ frame does have the same form as $F,$ i.e., that $%
F^{\prime }=\mathbf{E}^{\prime }+c\gamma _{5}\mathbf{B}^{\prime },$ where it
is interpreted that the LT of $\mathbf{E}$ and $\mathbf{B}$ are
\begin{eqnarray}
\mathbf{E}^{\prime } &=&(1+\gamma )/2)\mathbf{E+}\gamma \beta \gamma
_{0}(\gamma _{1}\cdot c\gamma _{5}\mathbf{B})+((\gamma -1)/2)\gamma
_{0}\gamma _{1}\mathbf{E}\gamma _{0}\gamma _{1},  \nonumber \\
c\gamma _{5}\mathbf{B}^{\prime } &=&((1+\gamma )/2)c\gamma _{5}\mathbf{B}%
+\gamma \beta F^{0k}\gamma _{k}\wedge \gamma _{1}+  \label{eb2} \\
&&((\gamma -1)/2)\gamma _{0}\gamma _{1}(c\gamma _{5}\mathbf{B)}\gamma
_{0}\gamma _{1}.  \nonumber
\end{eqnarray}
The relations (\ref{eb2}) are completely meaningless from the SR viewpoint
and they have nothing to do with the LT of a 4D quantity. In general, the LT
of some parts of a well-defined 4D quantity are not mathematically correct.
Hestenes in ''New Foundations for Classical Mechanics,'' $\left[ 3\right] $
p.625, declares: ''Considering the simplicity of the transformation law
(3.48) (our $F^{\prime }=RF\widetilde{R}$) for $F,$ it is obviously
preferable to treat the electromagnetic field $F=\mathbf{E}+c\gamma _{5}%
\mathbf{B}$ as a unit, rather than transform $\mathbf{E}$ and $\mathbf{B}$
separately by (3.51a,b) (our (\ref{eb2})).'' Our objection is that this is
not the question of the preferability but the question of the correctness.
The transformations (\ref{eb2}) are not less preferable than the
transformation $F^{\prime }=RF\widetilde{R}$ but the transformations (\ref
{eb2}) are, as we said, mathematically (and physically) incorrect. Thence
the same holds for the decomposition $F=\mathbf{E}+c\gamma _{5}\mathbf{B}$
and, more generally, for the space-time split of any well-defined 4D
physical quantity.

The similar result can be obtained with the passive LT. The passive LT
transform always the whole 4D quantity, basis and components, leaving the
whole quantity unchanged. This does not hold if $F$ is decomposed into the
relative vectors $\mathbf{E}$ and $\mathbf{B}$. Namely under the passive LT
it must hold that $F=(1/2)F^{\mu \nu }\gamma _{\mu }\wedge \gamma _{\nu
}=(1/2)F^{\mu ^{\prime }\nu ^{\prime }}\gamma _{\mu ^{\prime }}\wedge \gamma
_{\nu ^{\prime }}$ (the primed quantities are the Lorentz transforms of the
unprimed ones), which will not be fulfilled if $F$ is written in terms of $%
\mathbf{E}$ and $\mathbf{B}$, i.e., as $F=\mathbf{E}+c\gamma _{5}\mathbf{B.}$
The relative vectors $\mathbf{E}$ and $\mathbf{B}$ will transform under the
passive LT (i.e., when a basis is introduced) to another quantities, the
relative vectors $\mathbf{E}^{\prime }$ and $\mathbf{B}^{\prime }$ \emph{%
different} than $\mathbf{E}$ and $\mathbf{B}$ (the components and the basis
are changed; these transformations correspond to the transformations of the
usual 3D vectors $\mathbf{E}$ and $\mathbf{B,}$ e.g., $\left[ 11\right] ,$
eq. (11.149)). However the well-defined, geometric, 4D quantities, e.g., the
field bivector $F$ or \emph{1-vectors of electric and magnetic fields}, $E$
and $B$ respectively, introduced below, \emph{remain unchanged} under the
passive LT (for $F$ see above and, e.g., for $E$ it holds that $E=E^{\mu
}\gamma _{\mu }=E^{\mu ^{\prime }}\gamma _{\mu ^{\prime }}$).

\emph{These results (both with the active and the passive LT) entail that
the transformations of relative vectors} $\mathbf{E}$ \emph{and} $\mathbf{B}$
\emph{(as parts of a well-defined 4D quantity) are not mathematically
correct, which means that} $\mathbf{E}$ \emph{and} $\mathbf{B}$ \emph{%
themselves are not correctly defined quantities from the SR viewpoint}.
(Therefore it is not true from the SR viewpoint that ($\left[ 11\right] $
Sec. 11.10): ''A purely electric or magnetic field in one coordinate system
will appear as a mixture of electric and magnetic fields in another
coordinate frame.''; or that ($\left[ 3\right] $, Handout 10 in \textit{%
Physical Applications of Geometric Algebra)}: ''Observers in relative motion
see different fields.'') This is very important since it shows that, in
contrast to the generally accepted opinion, \emph{the usual ME with }$%
\mathbf{E}$ \emph{and} $\mathbf{B}$ \emph{are not relativistically correct
and thus they are not equivalent to the relativistically correct FE with }$F$
(\ref{MEF}). The same conclusion is achieved in the invariant formulation of
SR with tensors $\left[ 5\right] ,$ see particularly Sec. 5.3. in $\left[
5\right] $. \bigskip

\textbf{IV. THE\ FORMULATION\ OF\ ELECTRODYNAMICS WITH\ }$E$\textbf{\ AND }$B%
\mathbf{\ \bigskip }$

It is clear from the above consideration that the electrodynamics cannot be
correctly described from the point of view of SR with such quantities ($%
\mathbf{E}$ and $\mathbf{B}$) which do not transform properly under the LT.
\medskip

\textbf{A. Field equations with 1-vectors }$E$ and $B$ \textbf{\medskip }

Therefore instead of to decompose $F$ into $\mathbf{E}$ and $\mathbf{B}$ (%
\ref{FB}), which are not well-defined quantities from the SR viewpoint, we
present \emph{the observer independent} decomposition of $F$ by using
well-defined quantities in the Clifford algebra defined on the 4D spacetime,
the vectors (grade-1) of electric $E$ and magnetic $B$ fields. We define
\begin{equation}
F=(1/c)E\wedge v+e_{5}B\cdot v.  \label{myF}
\end{equation}
Conversely the relations which determine $E$ and $B$ in terms of $F$ are
\begin{equation}
E=(1/c)F\cdot v,\quad e_{5}B=(1/c^{2})F\wedge v,\ B=-(1/c^{2})e_{5}(F\wedge
v),  \label{myEB}
\end{equation}
and it holds that $E\cdot v=B\cdot v=0$ (since $F$ is antisymmetric), $v$ is
the velocity (1-vector) of a family of observers who measures $E$ and $B$
fields. The relations (\ref{myF}) and (\ref{myEB}) establish the equivalence
of the standard Clifford algebra formulation of electrodynamics with the
field bivector $F$ and the formulation with the 1-vectors of electric $E$
and magnetic $B$ fields. It is worth noting that now the observers in
relative motion see the same field, e.g., the $E$ field in the $S$ frame is
the same as in the relatively moving $S^{\prime }$; there is no mixture of $%
E $ and $B$ fields in $S^{\prime }$. The FE (\ref{MEF}) can be written in
terms of 1-vectors $E$ and $B$ as
\begin{equation}
\partial ((1/c)E\wedge v+e_{5}B\cdot v)=j/\varepsilon _{0}c.  \label{deb}
\end{equation}
In some basis $\left\{ e_{\mu }\right\} $ the FE (\ref{deb}) can be written
as
\begin{eqnarray}
\partial _{\alpha }(\delta _{\quad \mu \nu }^{\alpha \beta }v^{\mu }E^{\nu
}-\varepsilon ^{\alpha \beta \mu \nu }v_{\mu }cB_{\nu })e_{\beta }
&=&-(j^{\beta }/\varepsilon _{0})e_{\beta },  \nonumber \\
\partial _{\alpha }(\delta _{\quad \mu \nu }^{\alpha \beta }v^{\mu }cB^{\nu
}+\varepsilon ^{\alpha \beta \mu \nu }v_{\mu }E_{\nu })e_{5}e_{\beta } &=&0,
\label{maeb}
\end{eqnarray}
where $E^{\alpha }$ and $B^{\alpha }$ are the basis components of the
electric and magnetic 1-vectors $E$ and $B$, and $\delta _{\quad \mu \nu
}^{\alpha \beta }=\delta _{\,\,\mu }^{\alpha }\delta _{\,\,\nu }^{\beta
}-\delta _{\,\,\nu }^{\alpha }\delta _{\,\mu }^{\beta }.$ The first equation
in (\ref{maeb}) (the equation with sources) emerges from $\partial \cdot
F=j/\varepsilon _{0}c$ and the second one (the source-free equation) is
obtained from $\partial \wedge F=0.$ We remark that (\ref{maeb}) follows
from (\ref{deb}) for those coordinations for which the basis 1-vectors $%
e_{\mu }$ are constant, e.g., the standard basis $\left\{ \gamma _{\mu
}\right\} $ (the EC). For a nonconstant basis, for example, when one uses
polar or spherical basis 1- vectors (and, e.g., the Einstein
synchronization) then one must also differentiate these nonconstant basis
1-vectors. Instead of to work with $F$- formulation (\ref{MEF}) one can
equivalently use the $E,B$-formulation with the FE (\ref{deb}), or in the $%
\left\{ e_{\mu }\right\} $ basis (\ref{maeb}). For the given sources $j$ one
could solve these equations and find the general solutions for $E$ and $B.$
(We note that the equivalent formulation of electrodynamics with tensors $%
E^{a}$ and $B^{a}$ is reported in $\left[ 5\right] ,$ while the component
form in the EC\ is given in $\left[ 8\right] ,$ $\left[ 12\right] $ and $%
\left[ 13\right] .$)\bigskip

\textbf{B. Comparison with the usual noncovariant approach with the
3-vectors }$\mathbf{E}$\textbf{\ and }$\mathbf{B\medskip }$

The comparison of this geometric approach with Clifford 1-vectors $E$ and $B$
and the usual noncovariant approach with the 3D vectors $\mathbf{E}$ and $%
\mathbf{B}$ is possible in the EC. If one considers the EC and takes that in
an IFR $\mathcal{R}$ the observers who measure the basis components $%
E^{\alpha }$ and $B^{\alpha }$ are at rest, i.e., $v^{\alpha }=(c,0,0,0)$,
then in $\mathcal{R}$ $E^{0}=B^{0}=0$. Note that we can select a particular
- but otherwise arbitrary - IFR as the $\mathcal{R}$ frame, to which we
shall refer as the frame of our ''fiducial'' observers (see $\left[
11\right] $). In this frame of fiducial observers one can derive from the FE
in the $\left\{ \gamma _{\mu }\right\} $ basis (\ref{maeb}) the FE which
contain only the space parts $E^{i}$ and $B^{i}$ of $E^{\alpha }$ and $%
B^{\alpha }$, e.g., from the first FE in (\ref{maeb}) (the 1-vector part of (%
\ref{deb})) one easily finds
\begin{equation}
(\partial _{0}E^{i}-c\varepsilon ^{ijk0}\partial
_{j}B_{k}+j^{i}/c\varepsilon _{0})\gamma _{i}-(\partial
_{k}E^{k}-j^{0}/c\varepsilon _{0})\gamma _{0}=0,  \label{gam}
\end{equation}
and the second FE in (\ref{maeb}) (the trivector (pseudovector) part of (\ref
{deb})) yields
\begin{equation}
(c\partial _{0}B^{i}+\varepsilon ^{ijk0}\partial _{j}E_{k})\gamma _{5}\gamma
_{i}-(c\partial _{k}B^{k})\gamma _{5}\gamma _{0}=0.  \label{fag}
\end{equation}
The relations (\ref{gam}) and (\ref{fag}) are coordinate-based geometric
equations in the $\mathcal{R}$ frame and cannot be further simplified as
geometric equations. In the equation (\ref{gam}) one recognizes \emph{two}
\emph{ME} in the \emph{component form}, the Amp\`{e}re-Maxwell law $%
\varepsilon ^{ijk0}\partial _{j}B_{k}=j^{i}/c^{2}\varepsilon
_{0}+(1/c)\partial _{0}E^{i}$ (the first bracket, with $\gamma _{i}$) and
the Gauss law for the electric field $\partial _{i}E^{i}=j^{0}/\varepsilon
_{0}c$ (the second bracket, with $\gamma _{0}$). Similarly from (\ref{fag})
we find the \emph{component form} of another \emph{two ME}, Faraday's law $%
\varepsilon ^{ijk0}\partial _{j}E_{k}=c\partial _{0}B^{i}$ and the Gauss law
for the magnetic field $\partial _{k}B^{k}=0.$ It has to be noted that
neither in $\mathcal{R}$ there is a complete mathematical equivalence of
1-vectors $E$ and $B$ and the 3D $\mathbf{E}$ and $\mathbf{B.}$ Although the
components of the 3D $\mathbf{E}$ and $\mathbf{B}$ and the components of
1-vectors $E$ and $B$ are the same in $\mathcal{R}$ and in the EC the usual
3D vectors and the 1-vectors, when taken as geometric quantities, i.e.,
together with their bases, are mathematically different quantities; the
first quantities ($\mathbf{E}$ and $\mathbf{B}$) are defined on the 3D space
while the second ones ($E$ and $B$) are defined on the 4D spacetime. This
means that from (\ref{gam}) and (\ref{fag}) one cannot derive ME with the
usual 3D vectors $\mathbf{E}$ and $\mathbf{B,}$ but only the component form
of these equations in which $E^{i}$ and $B^{i}$ are the spatial components
of the 1-vectors $E$ and $B$ (the 4D quantities)$.$ From this consideration
one concludes that all the results obtained in the previous treatments from
the usual ME with the 3D $\mathbf{E}$ and $\mathbf{B}$ remain valid in the
formulation with the 1-vectors $E$ and $B$ if physical phenomena are
considered only in one IFR. Namely the selected IFR can be chosen to be the $%
\mathcal{R}$ frame. Then in $\mathcal{R}$, as explained above, the
coordinate-based geometric equations (\ref{gam}) and (\ref{fag}) can be
reduced to the equations containing only the components, the four ME in the
component form. Thus for observers who are at rest in $\mathcal{R}$ ($%
v^{\alpha }=(c,0,0,0)$) the components of the 3D $\mathbf{E}$ and $\mathbf{B}
$ (remember that $\mathbf{E}$ and $\mathbf{B}$ are not well defined
quantities from the SR viewpoint) can be simply replaced by the space
components of the 1-vectors $E$ and $B$ in the $\left\{ \gamma _{\mu
}\right\} $ basis. It has to be noted that just such observers are usually
considered in the conventional formulation with the 3D $\mathbf{E}$ and $%
\mathbf{B.}$ However, the situation is quite different when some physical
phenomena are considered from two relatively moving IFRs, say $S$ and $%
S^{\prime },$ for example, in the experiments that test SR. One of the
frames, say the $S$ frame, can be selected to be the $\mathcal{R}$ frame.
Now, even in the $\mathcal{R}$ frame we cannot simply use the four ME in the
component form, but from the outset we have to deal with the
coordinate-based geometric equations (\ref{maeb}), that is, (\ref{gam}) and (%
\ref{fag}) in $\mathcal{R}$, every of which contains two ME (the component
form) together. This means that it is not correct from the SR viewpoint to
investigate in two relatively moving IFRs, e.g., the Faraday law taken
alone. The Faraday law is always combined with the Gauss law for the
magnetic field in one unique law. They together form the relativistically
correct geometric equation (\ref{maeb}) (the second one), i.e., the equation
(\ref{fag}) in the $\mathcal{R}$ frame. This is an important difference
between our approach and all previous treatments and it will be examined in
more detail in our future work. The dependence of the FE (\ref{maeb}) on $v$
reflects the arbitrariness in the selection of the frame $\mathcal{R}$ but
at the same time it makes the equations (\ref{maeb}) independent of that
choice. The frame $\mathcal{R}$ can be selected at our disposal, which
proves that we don't have a kind of the ''preferred'' frame theory. We see
that the relativistically correct fields $E$ and $B$ and the new FE (\ref
{deb}) and (\ref{maeb}) do not have the same physical interpretation as the
usual, but relativistically incorrect, 3D fields $\mathbf{E}$ and $\mathbf{B}
$ and the usual 3D ME except in the frame $\mathcal{R}$ of the fiducial
observers in which $E^{0}=B^{0}=0$.

Furthermore an important general conclusion about the nonrelativistic
physics can be drawn from the above consideration. Namely if our living
arena is the 4D spacetime then only the geometric 4D quantities ($F,E,B,x,..$%
) can be correctly defined and can have an independent reality. The 3D
quantities from the nonrelativistic physics, both classical and quantum,
e.g., ($\mathbf{E,B,}t,\mathbf{x,}E$(energy)$,\mathbf{p}$(momentum),
Schr\"{o}dinger's $\psi ,\widehat{\mathbf{p}}$(3D momentum operator),...),
do not exist by themselves, and cannot have an independent reality in the 4D
spacetime. In order that such 3D quantities can be considered as properly
defined in the 4D spacetime they have to be (as in our consideration with $%
\mathbf{E}$ and $\mathbf{B}$) defined as parts of the corresponding
geometric 4D quantities taken in a specific frame and, usually, in the
EC.\medskip

\textbf{C. The stress-energy vector} $T(v)$ \textbf{and the energy-momentum
tensor} $T^{\mu \nu }$ \textbf{in the} $E,B$ \textbf{formulation}$\medskip $

It is shown in Sec. III that the stress-energy vector $T(v)$ for the
electromagnetic field is defined in the $F$ formulation by the relation (\ref
{ten}). Then in the standard basis $\left\{ \gamma _{\mu }\right\} $ we can
write the stress-energy vectors $T^{\mu }$ as $T^{\mu }=T(\gamma ^{\mu
})=(-\varepsilon _{0}/2)F\gamma ^{\mu }F.$ The components of the $T^{\mu }$
represent the energy-momentum tensor $T^{\mu \nu }$ in the $\left\{ \gamma
_{\mu }\right\} $ basis $T^{\mu \nu }=T^{\mu }\cdot \gamma ^{\nu
}=(-\varepsilon _{0}/2)\left\langle F\gamma ^{\mu }F\gamma ^{\nu
}\right\rangle $ (remember that $\left\langle A\right\rangle $ denotes the
scalar (grade-$0$) part of $A$), which reduces to familiar tensor form by
writing $F$ as $F=(1/2)F^{\mu \nu }\gamma _{\mu }\wedge \gamma _{\nu }$,

\noindent $T^{\mu \nu }=\varepsilon _{0}\left[ F^{\mu \alpha }g_{\alpha
\beta }F^{\beta \nu }+(1/4)F^{\alpha \beta }F_{\alpha \beta }g^{\mu \nu
}\right] .$

In the usual Clifford algebra aproach, e.g., $\left[ 1,2\right] $, one again
makes the space-time split and considers the energy-momentum density in the $%
\gamma _{0}$-system, $T^{0}=T(\gamma ^{0})=T(\gamma _{0});$ the split $%
T^{0}\gamma ^{0}=T^{0}\gamma _{0}=T^{00}+\mathbf{T}^{0},$ separates $T^{0}$
into an energy density $T^{00}=T^{0}\cdot \gamma ^{0}$ and a momentum
density $\mathbf{T}^{0}=T^{0}\wedge \gamma ^{0}.$ From the expression for $%
T^{\mu }$ and the relations (\ref{FB}) one finds the familiar results for
the energy density $T^{00}=(\varepsilon _{0}/2)(\mathbf{E}^{2}+c^{2}\mathbf{B%
}^{2})$ and the Poyinting vector $\mathbf{T}^{0}=\varepsilon _{0}(\mathbf{%
E\times }c\mathbf{B),}$ where the commutator product $A\times B$ is defined
as $A\times B\equiv (1/2)(AB-BA)$.

However, as we have already explained, the space-time split is not
relativistically correct procedure and the relative vectors $\mathbf{E}$ and
$\mathbf{B}$ are not well-defined quantities on the 4D spacetime.

Therefore we express the stress-energy vector $T(v)$ in terms of 1-vectors $%
E $ and $B.$ Inserting the relation for $F$ (\ref{myF}) into (\ref{ste}) we
express $T(v)$ by means of $E$ and $B$ ($v$ is again the velocity (1-vector)
of a family of observers who measures $E$ and $B$ fields) in a compact and
perspicuous form as
\begin{equation}
T(v)=(-\varepsilon _{0}/2c)(E^{2}+c^{2}B^{2})v+\varepsilon _{0}e_{5}\left[
\left( E\wedge B\right) \wedge v\right] .  \label{tv}
\end{equation}
$T(v)$ defined by (\ref{tv}) is frame and coordination independent quantity
and it is again written as a sum of the $v-\parallel $ and the $v-\perp $
parts. Thence the first term in (\ref{tv}) ($v-\parallel $ ) yields the
energy density $U$ as $U=(-\varepsilon _{0}/2)(E^{2}+c^{2}B^{2}),$ and the
second term ($v-\perp $) is $(1/c)$ of the Poynting vector $S=c\varepsilon
_{0}e_{5}\left[ \left( E\wedge B\right) \wedge v\right] .$ The observer
independent momentum density $g$ is defined as before $g=(1/c^{2})S$ and the
angular-momentum density is $M=(1/c)T(v)\wedge x=(U/c^{2})v\wedge x+g\wedge
x,$ where $T(v)$ is given by the relation (\ref{tv}).

All these quantities can be written in some basis $\left\{ e_{\mu }\right\} $
as CBGQs. Thus $T(v)$ (\ref{tv}) becomes
\begin{equation}
T(v)=(-\varepsilon _{0}/2c)(E^{\alpha }E_{\alpha }+c^{2}B^{\alpha }B_{\alpha
})v^{\lambda }e_{\lambda }+\varepsilon _{0}\widetilde{\varepsilon }_{\
\alpha \beta }^{\lambda }E^{\alpha }B^{\beta }e_{\lambda },  \label{tieb}
\end{equation}
where $\widetilde{\varepsilon }_{\lambda \alpha \beta }=\varepsilon _{\rho
\lambda \alpha \beta }v^{\rho }$ is the totally skew-symmetric Levi-Civita
pseudotensor induced on the hypersurface orthogonal to $v$. The energy
density $U$ ($U=v\cdot T(v)/c$) in the $\left\{ e_{\mu }\right\} $ basis is
determined by the first term in (\ref{tieb}) $U=(-\varepsilon
_{0}/2)(E^{\alpha }E_{\alpha }+c^{2}B^{\alpha }B_{\alpha }),$ and the
Poynting vector $S$ in the $\left\{ e_{\mu }\right\} $ basis is determined
by the second term in (\ref{tieb}) as $S=c\varepsilon _{0}\widetilde{%
\varepsilon }_{\ \alpha \beta }^{\lambda }E^{\alpha }B^{\beta }e_{\lambda }.$
Of course from (\ref{tieb}) one can easily find $g$ and $M$ in the $\left\{
e_{\mu }\right\} $ basis.

Although we don't need the energy-momentum tensor $T^{\mu \nu }$ (which is
defined in the $\left\{ e_{\mu }\right\} $ basis as $T^{\mu \nu }=T^{\mu
}\cdot e^{\nu }=(-\varepsilon _{0}/2)\left\langle Fe^{\mu }Fe^{\nu
}\right\rangle $ ) we quote here $T^{\mu \nu }$ expressed in terms of
components of 1-vectors $E$ and $B$ in some basis $\left\{ e_{\mu }\right\} $
as
\begin{eqnarray}
T^{\mu \nu } &=&\varepsilon _{0}[(g^{\mu \nu }/2-v^{\mu }v^{\nu
}/c^{2})(E^{\alpha }E_{\alpha }+c^{2}B^{\alpha }B_{\alpha })-(E^{\mu }E^{\nu
}+c^{2}B^{\mu }B^{\nu })+  \nonumber \\
&&(1/c)(\varepsilon ^{\mu \alpha \beta \lambda }v_{\lambda }v^{\nu
}+\varepsilon ^{\nu \alpha \beta \lambda }v_{\lambda }v^{\mu })B_{\alpha
}E_{\beta }],  \label{t3}
\end{eqnarray}
(see also $\left[ 12\right] $ and the first paper in $\left[ 8\right] $ for
the component form $T^{\mu \nu }$ in the EC). It has to be emphasized once
again that, in contrast to all earlier definitions including the Riesz
definition $\left[ 14\right] $ for the energy-momentum tensor $T^{\mu \nu },$
our definitions of $T(v),$ $U,$ $S,$ $g$ and $M$ are the definitions of
invariant quantities, i.e., frame and coordination independent quantities.

One can compare these expressions with familiar ones from the 3D space
considering our definitions in the standard basis $\left\{ \gamma _{\mu
}\right\} $ and in the $\mathcal{R}$ frame, the frame of our fiducial
observers, where $v^{\alpha }=(c,\mathbf{0})$, and consequently $%
E^{0}=B^{0}=0$. Then $U$ takes the familiar form $U=(-\varepsilon
_{0}/2)(E^{i}E_{i}+c^{2}B^{i}B_{i}),\ i=1,2,3.$ Similarly, in $\mathcal{R}$,
the Poynting vector becomes the familiar expression $S=\varepsilon
_{0}c^{2}\varepsilon _{0\ jk}^{\ i}E^{j}B^{k}\gamma _{i},$ $i,j,k=1,2,3,$
whence one also easily finds $g$ and $M$ in $\mathcal{R}$. Notice that all
quantities in these expressions are well-defined quantities on the 4D
spacetime. This again nicely illustrates our main idea that 3D quantities
don't exist by themselves but only as well-defined 4D quantities taken in a
particular - but otherwise arbitrary - IFR, here the $\mathcal{R}$ frame
(with fiducial observers).\medskip

\textbf{D. The Lorentz force in the }$E,B$ \textbf{formulation\medskip }

In the usual Clifford algebra approach to SR, e.g., $\left[ 1-3\right] $,
one makes the space-time split and writes the Lorentz force $K$ in the Pauli
algebra of $\gamma _{0}$. Since this procedure is observer dependent we
express $K$ in an \emph{observer independent way} using 1-vectors $E$ and $B$
as
\begin{equation}
K=(q/c)\left[ (1/c)E\wedge v+e_{5}B\cdot v\right] \cdot u.  \label{KEB}
\end{equation}
In the general case when charge and observer have distinct worldlines the
Lorentz force $K$ (\ref{KEB}) can be written as a sum of the $v-\perp $ part
$K_{\perp }$ and the $v-\parallel $ part $K_{\parallel },$ $K=K_{\perp
}+K_{\parallel },$ where
\begin{equation}
K_{\perp }=(q/c^{2})(v\cdot u)E+(q/c)(e_{5}B\cdot v)\cdot u,  \label{Kaok}
\end{equation}
\begin{equation}
K_{\parallel }=(-q/c^{2})(E\cdot u)v,  \label{kapa}
\end{equation}
respectively. This is an observer independent decomposition of the Lorentz
force $K.$ It can be easily verified that $K_{\perp }\cdot v=0$ and $%
K_{\parallel }\wedge v=0.$ Both parts can be written in some basis $\left\{
e_{\mu }\right\} $ as CBGQs $K_{\perp }=(q/c^{2})(v^{\mu }u_{\mu })E^{\nu
}e_{\nu }+(q/c)\widetilde{\varepsilon }_{\ \nu \rho }^{\mu }u^{\nu }B^{\rho
}e_{\mu },$ where, as already said, $\widetilde{\varepsilon }_{\mu \nu \rho
}\equiv \varepsilon _{\lambda \mu \nu \rho }v^{\lambda }$, and

\noindent $K_{\parallel }=(-q/c^{2})(E^{\mu }u_{\mu })v^{\nu }e_{\nu }.$
Speaking in terms of the prerelativistic notions one can say that in the
approach with the 1-vectors $E$ and $B$ $K_{\perp }$ plays the role of the
usual Lorentz force lying on the 3D hypersurface orthogonal to $v$, while $%
K_{\parallel }$ is related to the work done by the field on the charge.
However \emph{in our invariant formulation of SR only both components
together, (\ref{Kaok}) and (\ref{kapa}), do have physical meaning and they
define the Lorentz force both in the theory and in experiments. }

Let us consider a special case, the Lorentz force acting on a charge as
measured by a comoving observer ($v=u)$. Then from the definition of $K$ and
of $E$ (\ref{myEB}) one finds that $K=(q/c)F\cdot v=qE.$ Thus the Lorentz
force ascribed by an observer comoving with a charge is \emph{purely electric%
}. The relation
\begin{equation}
E\equiv \lim_{q\rightarrow 0}K/q  \label{elf}
\end{equation}
defines the electric field $E$ (1-vector) as the ratio of the measured force
$K$ (1-vector) on a stationary charge to the charge in the limit when the
charge goes to zero. Having $E$ so defined the charge can be given a
convenient uniform velocity $u\neq v$ from which the magnetic field $B$
(1-vector) is defined from the limit
\begin{equation}
\left[ (1/c)E\wedge v+e_{5}B\cdot v\right] \cdot u/c\equiv
\lim_{q\rightarrow 0}K/q.  \label{bef}
\end{equation}

When \emph{the complete} $K$ ((\ref{KEB}), or the sum of (\ref{Kaok}) and (%
\ref{kapa})), is known we can solve the equation of motion, Newton's second
law, written as
\begin{equation}
(q/c)\left[ (1/c)E\wedge v+e_{5}B\cdot v\right] \cdot u=m(u\cdot \partial )u,
\label{eqmot}
\end{equation}
(the Lorentz force $K$ on the l.h.s. of (\ref{eqmot}) can be also written as
the sum of $K_{\perp }$ (\ref{Kaok}) and $K_{\parallel }$ (\ref{kapa})), $%
u\cdot \partial $ is the directional derivative and $(u\cdot \partial )u$
defines the acceleration of the particle. From Newton's second law (\ref
{eqmot}) one obtains the force $K.$ From $K$ and the definitions (\ref{elf})
and (\ref{bef}) one determines the 1-vectors $E$ and $B.$ This form of the
equation of motion differs from the usual Clifford algebra approach to SR,
e.g., $\left[ 1-3\right] ,$ but it agrees with the tensor formulation in
general relativity, see, e.g., $\left[ 15\right] $. Notice, however, that
such form (\ref{eqmot}) is used here in SR since in this invariant
formulation of SR one can use different coordinations of an IFR. Thence, in
general, the derivatives of the nonconstant basis vectors must be also taken
into account, e.g., if one uses the Einstein synchronization and polar or
spherical spatial coordinate basis. Therefore the equation of motion has to
be written in the form of the equation (\ref{eqmot}).

These expressions for $K$ can be compared with familiar ones from the 3D
space considering our results in the standard basis $\left\{ \gamma _{\mu
}\right\} $ and in the $\mathcal{R}$ frame, the frame of our fiducial
observers. Then $K_{\perp }^{0}e_{0}=0,$ and, e.g., $K_{\perp
}^{1}e_{1}=qE^{1}e_{1}+q(u^{2}B^{3}-u^{3}B^{2})e_{1}.$ Further $K_{\parallel
}^{0}e_{0}=(-q/c)(E^{i}u_{i})e_{0},$ and all $K_{\parallel }^{i}=0,$ $%
i=1,2,3.$ We see that in the $\mathcal{R}$ frame the whole $K$ (\ref{KEB})
does have $K^{i}=K_{\perp }^{i},$ $i=1,2,3$ and, as seen from the above
expressions, equal to the usual 3D expression for the Lorentz force, while $%
K^{0}=K_{\parallel }^{0},$ and represents the usual 3D expression for the
work done by the field on the charge. Then in the $\mathcal{R}$ frame the
definitions of $E$ (\ref{elf}) and $B$ (\ref{bef}) become
\begin{equation}
E^{i}e_{i}\equiv \lim_{q\rightarrow 0}(K_{\perp }^{i}/q)e_{i},\quad
E^{0}e_{0}=0  \label{elf1}
\end{equation}
and
\begin{equation}
(E^{i}+\varepsilon ^{0kji}u_{k}B_{j})e_{i}\equiv \lim_{q\rightarrow
0}(K_{\perp }^{i}/q)e_{i},\quad B^{0}e_{0}=0.  \label{bef1}
\end{equation}
The expressions (\ref{elf1}) and (\ref{bef1}) correspond to the usual
definitions of the 3D $\mathbf{E}$ and $\mathbf{B}$ in terms of the 3D
expression for the Lorentz force. However it has to be noted that $e_{i}$
and $e_{0}$ in (\ref{elf1}) and (\ref{bef1}) are the 1-vectors (defined on
the 4D spacetime) while the 3D $\mathbf{E}$ and $\mathbf{B}$ and the usual
3D Lorentz force are defined on the 3D space.\bigskip

\textbf{V. THE\ FORMULATION\ OF\ ELECTRODYNAMICS WITH\ THE REAL }$\Psi
=E-ce_{5}B$ \textbf{\bigskip }

In this section we consider the formulation of electrodynamics with Clifford
aggregates $\Psi $ and $\widetilde{\Psi }$ and show that it is equivalent to
those with $F$ and with $E$ and $B.$
\begin{eqnarray}
\Psi &=&E-ce_{5}B,\quad \widetilde{\Psi }=E+ce_{5}B,  \nonumber \\
E &=&(1/2)(\Psi +\widetilde{\Psi }),\quad B=(1/2c)e_{5}(\Psi -\widetilde{%
\Psi }).  \label{pse}
\end{eqnarray}
In contrast to the usual decomposition of $F$ into the relative vectors $%
\mathbf{E}$ and $\mathbf{B}$ $\left[ 1-3\right] $ \emph{the multivectors} $%
\Psi $ \emph{and} $\widetilde{\Psi }$ \emph{defined by} (\ref{pse}) \emph{%
are frame and coordination independent quantities.}

Of course it is easy to find how the $F$ formulation and the $\Psi $
formulation are connected
\begin{eqnarray}
\Psi &=&(-1/c)(vF),\ \widetilde{\Psi }=(-1/c)(\widetilde{F}v)=(1/c)(Fv),
\nonumber \\
F &=&(-1/c)(v\Psi ),\ \widetilde{F}=(-1/c)(\widetilde{\Psi }v).  \label{efps}
\end{eqnarray}
Then we can write the FE (\ref{MEF}) in terms of the multivector $\Psi $ in
a simple form
\begin{eqnarray}
\partial (v\Psi ) &=&-j/\varepsilon _{0},  \nonumber \\
\partial \cdot (v\Psi ) &=&-j/\varepsilon _{0},\quad \partial \wedge (v\Psi
)=0.  \label{Mepsi}
\end{eqnarray}
The multivector $\Psi $ is not a homogeneous multivector but a mixed-grade
multivector; it is the sum of an 1-vector and a 3-vector (pseudovector).
However $v\Psi $ is a bivector (it determines $F$), see (\ref{efps}), and $%
v\Psi =v\cdot \Psi +v\wedge \Psi =-cv\cdot e_{5}B+v\wedge E.$

The equation (\ref{Mepsi}) can be written in some basis $\left\{ e_{\mu
}\right\} $ (in which the basis 1-vectors $e_{\mu }$ are constant) with the
CBGQs
\begin{equation}
\partial _{\alpha }(\delta _{\quad \mu \nu }^{\alpha \beta
}-e_{5}\varepsilon _{\quad \mu \nu }^{\alpha \beta })v^{\mu }\Psi ^{\nu
}e_{\beta }=-(j^{\beta }/\varepsilon _{0})e_{\beta },  \label{psme}
\end{equation}
where $\Psi ^{\alpha }e_{\alpha }=(E^{\alpha }-ce_{5}B^{\alpha })e_{\alpha }$%
. Using this last relation it can be seen that the equation (\ref{psme})
contains both equations with $E$ and $B$ from (\ref{maeb}). The equation (%
\ref{psme}) can be also written as
\begin{eqnarray}
((\Gamma ^{\alpha })_{\ \mu }^{\beta }\partial _{\alpha }\Psi ^{\mu
})e_{\beta } &=&-(j^{\beta }/\varepsilon _{0})e_{\beta },  \nonumber \\
(\Gamma ^{\alpha })_{\ \mu }^{\beta } &=&\delta _{\quad \rho \nu }^{\alpha
\beta }v^{\rho }g_{\ \mu }^{\nu }+e_{5}\varepsilon _{\quad \mu \nu }^{\alpha
\beta }v^{\nu }.  \label{gaps}
\end{eqnarray}
In the case that $j^{\beta }=0$ the equation (\ref{gaps}) becomes
\begin{equation}
((\Gamma ^{\alpha })_{\ \mu }^{\beta }\partial _{\alpha }\Psi ^{\mu
})e_{\beta }=0.  \label{gam1}
\end{equation}
It has to be emphasized that the equation (\ref{gam1}) is a real one;
Clifford algebra is developed over the field of the real numbers. Further it
is not the component form in the EC but it is a coordinate-based geometric
equation since it is written using CBGQs. We see from (\ref{gam1}) that the
FE for the free electromagnetic field become Dirac-like relativistic wave
equation for the free photon. Really (\ref{gam1}) is the one-photon quantum
equation when $\Psi $ is interpreted as the one-photon wave function and
when the continuity equation is introduced as for the complex $\Psi $ in $%
\left[ 13\right] $ (the component form in the EC). However since $\Psi $ is
real there is no need for the probabilistic interpretation of $\Psi $! This
will be discussed elsewhere.

Let us also write the stress-energy vector $T(v)$ in terms of $\Psi ,$ and
as a sum of the $v-\parallel $ and the $v-\perp $ parts. Then
\begin{equation}
T(v)=-(\varepsilon _{0}/4c)\left( \Psi \cdot \Psi +\widetilde{\Psi }\cdot
\widetilde{\Psi }\right) v+(\varepsilon _{0}/4c)\left( \Psi \cdot \Psi -%
\widetilde{\Psi }\cdot \widetilde{\Psi }\right) \cdot v.  \label{tps}
\end{equation}
Hence $U=-(\varepsilon _{0}/4)\left( \Psi \cdot \Psi +\widetilde{\Psi }\cdot
\widetilde{\Psi }\right) ,$ and the Poynting vector is

\noindent $S=(\varepsilon _{0}/4)\left( \Psi \cdot \Psi -\widetilde{\Psi }%
\cdot \widetilde{\Psi }\right) \cdot v.$ From (\ref{tps}) one also finds the
observer independent expressions for $g$ and $M$ in terms of $\Psi .$

All these quantities can be written in some basis $\left\{ e_{\mu }\right\} $
as CBGQs but we will not do it here.

Only we write the energy-momentum tensor $T^{\mu \nu }$ in terms of
components of $\Psi $ in some basis $\left\{ e_{\mu }\right\} $ as
\begin{eqnarray}
T^{\mu \nu } &=&\varepsilon _{0}[(g^{\mu \nu }/2-v^{\mu }v^{\nu }/c^{2})%
\widetilde{\Psi }^{\alpha }\Psi _{\alpha }-(1/2)(\widetilde{\Psi }^{\mu
}\Psi ^{\nu }+\Psi ^{\mu }\widetilde{\Psi }^{\nu })-  \nonumber \\
&&(e_{5}/2c)(\varepsilon ^{\mu \alpha \beta \lambda }v^{\nu }+\varepsilon
^{\nu \alpha \beta \lambda }v^{\mu })\widetilde{\Psi }_{\alpha }\Psi _{\beta
}v_{\lambda }].  \label{pst1}
\end{eqnarray}

The Lorentz force $K$ is given in the real $\Psi $ formulation as
\begin{eqnarray}
K &=&(q/c^{2})\left[ u\cdot (v\cdot \Psi +v\wedge \Psi )\right] =  \nonumber
\\
&&(q/c^{2})\left[ u\cdot (v\cdot \Psi )+(u\cdot v)\Psi -v\wedge (u\cdot \Psi
)\right] .  \label{klps}
\end{eqnarray}
\bigskip

\textbf{VI. THE\ FORMULATION\ OF\ ELECTRODYNAMICS WITH\ THE\ COMPLEX }$\Psi
=E-icB$ \textbf{\bigskip }

Sometimes it will be useful to work with the complex $\Psi $; Clifford
algebra is developed over the field of the complex numbers. Then
\begin{eqnarray}
\Psi &=&E-icB,\quad \overline{\Psi }=E+icB,  \nonumber \\
E &=&(1/2)(\Psi +\overline{\Psi }),\quad B=(i/2c)(\Psi -\overline{\Psi }),
\label{pkom}
\end{eqnarray}
$i$ is the unit imaginary. $\overline{\Psi }$ is the complex reversion of $%
\Psi $. In contrast to the real $\Psi $, which is a multivector of a mixed
grade (the sum of grade-1 and grade-3 multivectors), the complex $\Psi $ is
a homogeneous, grade-1, multivector. This fact facilitates the calculation
in some cases, particularly when one considers the relativistic quantum
mechanics. Notice also that now it holds that $v\cdot \Psi =v\cdot \overline{%
\Psi }=0.$

The $F$ formulation and the complex $\Psi $ formulation are connected by the
relations
\begin{eqnarray}
\Psi &=&(1/c)F\cdot v+(i/c)e_{5}(F\wedge v),  \nonumber \\
F &=&(1/2c)(\Psi +\overline{\Psi })\wedge v+(i/2c)\left( e_{5}(\Psi -%
\overline{\Psi })\right) \cdot v.  \label{f}
\end{eqnarray}

Then we can use the second equation from (\ref{f}) and the FE (\ref{MEF}) to
write FE in terms of the complex 1-vector $\Psi $ as
\begin{equation}
\partial \cdot (v\wedge \Psi )-ie_{5}\left[ \partial \wedge (v\wedge \Psi
)\right] =-j/\varepsilon _{0}.  \label{mecp}
\end{equation}
This form of FE (in which $\overline{\Psi }$ does not appear) is achieved
separating vector and trivector parts and then combining them to eliminate $%
\overline{\Psi }$.

Of course FE (\ref{mecp}) can be written in some basis $\left\{ e_{\mu
}\right\} $ (with constant $e_{\mu }$) with the CBGQs as
\begin{equation}
\partial _{\alpha }(\delta _{\quad \mu \nu }^{\alpha \beta }-i\varepsilon
_{\quad \mu \nu }^{\alpha \beta })v^{\mu }\Psi ^{\nu }e_{\beta }=-(j^{\beta
}/\varepsilon _{0})e_{\beta }.  \label{meko}
\end{equation}
This relation is of the same form as the equation (\ref{psme}) but $e_{5}$
is replaced by $i$ and in (\ref{meko}) $\Psi $ is a complex 1-vector. Again
it can be seen that the equation (\ref{meko}) contains both equations with $E
$ and $B$ from (\ref{maeb}). Further the equation (\ref{meko}) can be
written in the same form as (\ref{gaps}) only in $(\Gamma ^{\alpha })_{\ \mu
}^{\beta }$ the pseudoscalar $e_{5}$ is replaced by $i$
\begin{eqnarray}
((\Gamma ^{\alpha })_{\ \mu }^{\beta }\partial _{\alpha }\Psi ^{\mu
})e_{\beta } &=&-(j^{\beta }/\varepsilon _{0})e_{\beta },  \nonumber \\
(\Gamma ^{\alpha })_{\ \mu }^{\beta } &=&\delta _{\quad \rho \nu }^{\alpha
\beta }v^{\rho }g_{\ \mu }^{\nu }+i\varepsilon _{\quad \mu \nu }^{\alpha
\beta }v^{\nu }.  \label{gako}
\end{eqnarray}
The equations (\ref{meko}) and (\ref{gako}) are of the same form as the
corresponding equations written in the tensor formulation with the complex $%
\Psi $ in $\left[ 5\right] $ (or in the component form in the EC in $\left[
8\right] $ and $\left[ 13\right] $). Again for $j^{\beta }=0$ we find from (%
\ref{gako}) that the FE for the free electromagnetic field become Dirac-like
relativistic wave equation for the free photon.

Furthermore $T(v)$ can be expressed by $\Psi $ and $\overline{\Psi }$ as a
sum of the $v-\parallel $ and the $v-\perp $ parts, which determine $U$ and $%
S.$ Thus $T(v)$ is given as
\begin{equation}
T(v)=-(\varepsilon _{0}/2c)\left( \Psi \cdot \overline{\Psi }\right)
v-i(\varepsilon _{0}/2c)e_{5}\left[ \Psi \wedge \overline{\Psi }\wedge
v\right] .  \label{kot}
\end{equation}
Hence $U=-(\varepsilon _{0}/2)\left( \Psi \cdot \overline{\Psi }\right) $
and $S=-i(\varepsilon _{0}/2)e_{5}\left[ \Psi \wedge \overline{\Psi }\wedge
v\right] .$ In some basis $\left\{ e_{\mu }\right\} $ the stress-energy
vector $T(v)$ can be written as the CBGQ
\[
T(v)=(-\varepsilon _{0}/2c)(\Psi ^{\alpha }\overline{\Psi }_{\alpha
})v^{\lambda }e_{\lambda }-i(\varepsilon _{0}/2c)\widetilde{\varepsilon }_{\
\alpha \beta }^{\lambda }\Psi ^{\alpha }\overline{\Psi }^{\beta }e_{\lambda
}.
\]
where $\widetilde{\varepsilon }_{\lambda \alpha \beta }=\varepsilon _{\rho
\lambda \alpha \beta }v^{\rho }.$

Similarly we can write the energy-momentum tensor $T^{\mu \nu }$ in terms of
components of the complex $\Psi $ in some basis $\left\{ e_{\mu }\right\} $
and it is the same as (\ref{pst1}) except that the pseudoscalar $e_{5}$ is
replaced by $i.$

The Lorentz force $K$ can be again written as a sum of the $v-\perp $ and
the $v-\parallel $ parts $K=K_{\perp }+K_{\parallel },$ where
\begin{eqnarray}
K_{\perp } &=&(q/2c^{2})\left[ (u\cdot v)(\Psi +\overline{\Psi })+i\left(
u\wedge v\wedge (\Psi -\overline{\Psi })\right) e_{5}\right] ,  \nonumber \\
K_{\parallel } &=&-(q/2c^{2})\left[ u\cdot (\Psi +\overline{\Psi })v\right] .
\label{kop}
\end{eqnarray}
Thus all four formulations are presented in geometric terms, i.e., with
invariant quantities and can be equivalently used in all calculations.
\bigskip

\textbf{VII. COMPARISON\ WITH\ EXPERIMENTS \bigskip }

It is shown in $\left[ 6\right] $ that the usual formulation of SR (which
deals with the observer dependent quantities, i.e., the Lorentz contraction,
the dilatation of time, the use of the 3D $\mathbf{E}$ and $\mathbf{B}$,
etc.,) shows only an ''apparent'' agreement (not the true one) with the
traditional and modern experiments, e.g., the Michelson-Morley type
experiments. On the contrary the invariant SR from $\left[ 5\right] $ (given
in terms of \emph{geometric quantities - tensors}) is shown in $\left[
6\right] $ to be in a \emph{complete agreement} with all considered
experiments. This entails that the same complete agreement holds also for
the formulations with \emph{geometric quantities - the Clifford numbers},
which are presented in this paper.

In addition we briefly discuss the Trouton-Noble experiment $\left[
16\right] $ (see also $\left[ 17\right] $). In the experiment they looked
for the turning motion of a charged parallel plate capacitor suspended at
rest in the frame of the earth in order to measure the earth's motion
through the ether. The explanations, which are given until now (see, e.g., $%
\left[ 18\right] $), for the null result of the experiments $\left[
16\right] $ ($\left[ 17\right] $) are not relativistically correct, since
they use quantities that are not well-defined in 4D spacetime; e.g., the
Lorentz contraction, the transformation equations for the usual 3D vectors $%
\mathbf{E}$ and $\mathbf{B}$ and for the torque as the 3D vector, the
nonelectromagnetic forces of undefined nature, etc.. In our approach the
explanation is very simple and natural; the energy density $U,$ then $g$ and
$M$ and the associated integral quantities are all invariant quantities,
which means that their values are the same in the rest frame of the
capacitor and in the moving frame. Thus if there is no torque (but now as a
geometric, invariant, 4D quantity) in the rest frame then the capacitor
cannot appear to be rotating in a uniformly moving frame.\bigskip

\textbf{VIII. DISCUSSION\ AND\ CONCLUSIONS\bigskip }

The usual Clifford algebra approach to the relativistic electrodynamics
deals with the space-time split and the relative vectors $\mathbf{E}$ and $%
\mathbf{B.}$ The investigation presented in this paper reveals that such
approach is not relativistically correct. The relative vectors are not only
observer dependent but their transformation law is meaningless from the SR
viewpoint; it has nothing to do with the Lorentz transformations of Clifford
numbers defined on the 4D spacetime. Here we employ quantities that are
independent of the reference frame and of the chosen coordination for that
frame. We have presented four equivalent formulations of electrodynamics by
means of the field bivector $F,$ the 1-vectors $E$ and $B$, the real
multivector $\Psi =E-ce_{5}B$ and the complex 1-vector $\Psi =E-ciB.$ All
four formulations are equivalent and they yield complete and consistent
descriptions of electromagnetic phenomena in terms of observer independent,
thus properly defined quantities on the 4D spacetime. These formulations are
not equivalent with the usual Maxwell formulation with the 3D vectors $%
\mathbf{E}$ and $\mathbf{B}$ except in $\mathcal{R}$, the frame of fiducial
observers ($v^{\alpha }=(c,\mathbf{0})$ and consequently $E^{0}=B^{0}=0$),
and when the Einstein coordination is used in $\mathcal{R}$. \emph{The new
observer independent FE }with $E$ and $B$ (\ref{deb}), with the real
multivector $\Psi $ (\ref{Mepsi}) and with the complex 1-vector $\Psi $ (\ref
{mecp}) are presented in this paper. Furthermore \emph{the new observer
independent expressions }for the stress-energy vector $T(v),$ the energy
density $U,$ the Poynting vector $S,$ the momentum density $g,$ the
angular-momentum density $M$ and the Lorentz force $K$ are given in all four
formulations. The field equations with the real $\Psi $ (\ref{gam1}) (and
the corresponding equation for the complex $\Psi $) for the free
electromagnetic field ($j=0$) look like Dirac relativistic wave equation for
the free photon; there is no need to perform the first quantization
procedure. Particularly it is important to note that in (\ref{gam1}) the $%
\Psi $ function is the real one. Hence we don't need the probabilistic
interpretation for such $\Psi $! The second quantization procedure, and the
whole quantum electrodynamics, will be simply constructed using geometric,
invariant, quantities $E$ and $B,$ the real or the complex $\Psi $, $T(v)$, $%
U$, $S$, $g$ and $M.$ Note that the standard covariant approaches to quantum
electrodynamics, e.g., $\left[ 19\right] $, deal with the component form (in
the specific, i.e., the Einstein coordination) of the electromagnetic
4-potential $A$ (thus requiring the gauge conditions too) instead of to use
the geometric quantities, the fields 1-vectors $E$ and $B,$ the real or
complex $\Psi $ as Clifford numbers (this work), or the tensors $E^{a},$ $%
B^{a},$ or $\Psi ^{a}$ (see, e.g., $\left[ 5\right] $). Furthermore the
standard covariant approaches employ the definitions of the field energy and
momentum, which are not well-defined from the relativity viewpoint. Namely
both the field energy and momentum are defined as integrals over the \emph{%
three-space}, that is, over the hypersurface $t=const.$ But the hypersurface
$t=const.$ in some reference frame $S$ cannot become (under the Lorentz
transformation) the hypersurface $t^{\prime }=const.$ in a relatively moving
reference frame $S^{\prime }.$ This is already examined for the classical
electrodynamics (the covariant formulation in the EC) by Rohrlich $\left[
20\right] $ and in the first paper in $\left[ 8\right] $. Here the local
conservation laws are directly derived from the FE and written in an
invariant way. The observer independent integral FE and the observer
independent global conservation laws (with the definitions of the invariant
field energy and momentum) will be treated elsewhere. Particularly it has to
be emphasized that the observer independent approach to the relativistic
electrodynamics that is presented in this paper is in a complete agreement
with existing experiments that test SR, which is not the case with the usual
approaches. Furthermore we note that all observer independent quantities
introduced here and the FE written in terms of them hold in the same form
both in the flat and curved spacetimes. The formalism presented here will be
the basis for the relativistically correct (without reference frames)
formulation of quantum electrodynamics and, more generally, of the quantum
field theory. \bigskip

\noindent \textbf{REFERENCES}\bigskip

\noindent $\left[ 1\right] $ D. Hestenes, \textit{Space-Time Algebra }%
(Gordon and Breach, New York, 1966); \textit{Space-Time Calculus; }available
at: http://modelingnts.la. asu.edu/evolution.html; \textit{New Foundations
for Classical Mechanics }(Kluwer Academic Publishers, Dordrecht, 1999) 2nd.
edn..

\noindent $\left[ 2\right] $ S. Gull, C. Doran, and A. Lasenby, in \textit{%
Clifford (Geometric) Algebras with Applications to Physics, Mathematics, and
Engineering,} W.E. Baylis, Ed. (Birkhauser, Boston, 1997), Chs. 6-8.; Found.
Phys. \textbf{23}, 1175 (1993); Found. Phys. \textbf{23}, 1239 (1993);
Found. Phys. \textbf{23}, 1295 (1993); Found. Phys. \textbf{23}, 1329
(1993); C. Doran, and A. Lasenby, \textit{Physical Applications of Geometric
Algebra,} available at: www.mrao.cam.ac.uk/\symbol{126}Clifford/

\noindent $\left[ 3\right] $ B. Jancewicz, \textit{Multivectors and Clifford
Algebra in Electrodynamics} (World Scientific, Singapore, 1989).

\noindent $\left[ 4\right] $ D. Hestenes and G. Sobczyk, \textit{Clifford
Algebra to Geometric Calculus }(Reidel, Dordrecht, 1984).

\noindent $\left[ 5\right] $ T. Ivezi\'{c}, Found. Phys. \textbf{8}, 1139
(2001).

\noindent $\left[ 6\right] $ T. Ivezi\'{c}, Found. Phys. Lett. \textbf{15},
27 (2002); physics/0103026; physics/0101091.

\noindent $\left[ 7\right] $ T. Ivezi\'{c}, Annales de la Fondation Louis de
Broglie \textbf{27}, 287 (2002).

\noindent $\left[ 8\right] $ T. Ivezi\'{c}, Found. Phys. Lett. \textbf{12},
105 (1999); Found. Phys. Lett. \textbf{12}, 507 (1999).

\noindent $\left[ 9\right] $ A. Einstein, Ann. Physik. \textbf{17}, 891
(1905), tr. by W. Perrett and G.B.

Jeffery, in \textit{The Principle of Relativity} (Dover, New York).

\noindent \ $\left[ 10\right] $. D.E. Fahnline, Am. J. Phys. \textbf{50},
818 (1982).

\noindent $\left[ 11\right] $ J.D. Jackson, \textit{Classical Electrodynamics%
} (Wiley, New York, 1977) 2nd edn..

\noindent $\left[ 12\right] $ H.N. N\'{u}\~{n}ez Y\'{e}pez, A.L. Salas
Brito, and C.A. Vargas, Revista Mexicana de F\'{i}sica \textbf{34}, 636
(1988).

\noindent $\left[ 13\right] $ S. Esposito, Found. Phys. \textbf{28}, 231
(1998).

\noindent $\left[ 14\right] $ M. Riesz, Dixi\'{e}me Congr\'{e}s des
Mathematicians Scandinaves, Copenhague, 1946, p.123.

\noindent $\left[ 15\right] $ R.M. Wald, \textit{General Relativity} (The
University of Chicago Press, Chicago, 1984).

\noindent $\left[ 16\right] $ F.T. Trouton and H.R. Noble, Philos. Trans. R.
Soc. London Ser. A \textbf{202}, 165 (1903).

\noindent $\left[ 17\right] $ H.C. Hayden, Rev. Sci. Instrum. \textbf{65},
788 (1994).

\noindent $\left[ 18\right] $ A.K. Singal, J. Phys. A: Math. Gen. \textbf{25}
1605 (1992); Am. J. Phys. \textbf{61}, 428 (1993); S. A. Teukolsky, Am. J.
Phys. \textbf{64}, 1104 (1996); O.D. Jefimenko, J. Phys. A: Math. Gen.
\textbf{32,} 3755 (1999).

\noindent $\left[ 19\right] $ J.D. Bjorken and S.D. Drell, \textit{%
Relativistic Quantum Field} (McGraw-Hill, New York, 1964); F. Mandl and G.
Shaw, \textit{Quantum Field Theory} (John Wiley \&Sons, New York, 1995); S.
Weinberg, \textit{The} \textit{Quantum Theory of Fields, Vol. I Foundations,
}(Cambridge University Press, Cambridge, 1995).

\noindent $\left[ 20\right] $ F. Rohrlich, Phys. Rev. D \textbf{25}, 3251
(1982).

\end{document}